\def\np{Nucl. Phys.}
\def\pl{Phys. Lett.}
\def\prl{Phys. Rev. Lett.}
\def\pr{Phys. Rev.}

\def\be{\begin{equation}}
\def\ee{\end{equation}}
\def\ba{\begin{eqnarray}}
\def\ea{\end{eqnarray}}
\documentstyle[12pt]{article}

\begin{document}
\begin{titlepage}
\title{{\bf The upper bound of the top mass and electroweak radiative
corrections}}
\begin{small}
\author{ {\bf Francesco Caravaglios }  \\
Scuola Normale superiore di Pisa and Infn sezione di Pisa \\}
\date{}
\end{small}
\maketitle

\begin{abstract}
We investigated the possibility of introducing sizeable
negative corrections to the
$\varepsilon_{N1}$ ($\delta \rho$) parameter without affecting
$\varepsilon_{N3}$.
We have found that a proper vector-like family of fermions can imply
 such corrections.
Differently from supersymmetry \cite{bcf}, this
can be realized without introducing light particles easily
observable at LEP II. Our example can be of particular interest if
no new particle is found at LEP II
and the $\varepsilon_{N1}$ value is found to be small compared to
the one expected in the case  of a large top mass.
\end{abstract}
\vspace{3cm}
\noindent
IFUP-TH 3/94 \\
January 1994

\vskip-22.5cm
\rightline{IFUP-TH}

\end{titlepage}

\section*{Introduction} In this paper, we study the radiative corrections of
possible extensions of the standard model, paying particular attention to the
$\rho$ parameter which has relevant implications to the upper bound on the top
mass. Precision tests are mainly  relevant for two reasons [1-13]:
 \begin{itemize}
\item[1)] they could give signals of new physics; \item[2)] their strong
dependence on the top mass parameter gives an upper bound on this parameter.
\end{itemize}
Essentially, there are two measurements that provide us with an upper limit on
the top mass: the direct and indirect \cite{Akundov}
 measurement of the $Z_0$ partial width into bottom and anti-bottom quarks
(in other words $\varepsilon_b$ \cite{abc});
and the ratio between the axial-vector coupling to leptons of the
$Z_0$ and $W^\pm$ gauge bosons ($\varepsilon_1$ or $\delta \rho$ \cite{ab}
\cite{abc}).
  The first one gives an upper limit ($95 \%$ CL)
on the top mass near above $210~GeV$ \cite{abc}
which can be violated only
if new physics affects the $Z_0$-bottom vertex.
This limit results from the present measurement of $\varepsilon_b$
with an error of about $5$ per mill, which is dominated by the
error on $\alpha_s(M_Z^2)$. It is unlikely that this error will be
lower than $4$ per mill. So, except for large displacement of $\varepsilon_b$
central value, this limit will not change much in the next future.
On the contrary, the measurement of $\varepsilon_1$ combined possibly with
a top mass measurement (for instance from the CDF experiment) could provide
interesting  new physics effects.

 The experimental observation of the relation \be \label{rho} \rho \simeq 1 \ee
suggests that the symmetry breaking mechanism involves only Higgs doublets and
that deviations from the above relation are obtained only through radiative
corrections. Several  authors have studied radiative corrections within and
beyond the standard model \cite{bcf} \cite{casalbuoni} \cite{tecni} showing all
possible consequence for the precision measurements of $\varepsilon_1$ both at
low and high energy (LEP) experiments. As a general  conclusion, within the
models studied in the past, it is difficult to weaken the top mass upper bound
and, on the contrary, it is easy to lower this upper bound. In supersymmetric
models there is a way to weaken this limit \cite{bcf}, if some charginos are in
the domain explorable at LEP II. Such particles would give, if just above
the threshold of production at LEP I, a reduction of the Z-width ,affecting the
$\varepsilon_{N1}$ parameter through $e_5$ and raising the top mass upper
limit.
The price to pay is to introduce an analogous effect on the
 $\varepsilon_{N3}$
parameter and the necessity of having some particles just above LEP I and
easily
observable at LEP II. The present experimental result for the
$\varepsilon_{N1}$
parameter is \be \label{exp}  \varepsilon_{N1}=1.6\pm 2.6 \times 10^{-3} \ee
where the experimental error on $\varepsilon_{N1}$ will be (probably) 1-1.5 per
mill at the end of \hbox{LEP I} experiment. As an example, if the central value
of this measure is not raised , we will be likely to have in the next future a
statistically    significant  evidence for a  negative  contribution to
$\varepsilon_{N1}$, in particular if a heavy top shows up ($M_{top} > 150~(175)
GeV$ implies $\varepsilon_{N1} > 3.5~(5.8) \times 10^{-3}$ with a light Higgs).

In addition, if no new particle shows up at the LEP II phase\footnote{ In other
words if no $e_5$ contribution can explain the $\varepsilon_{N1}$ value
\cite{bcf}.} we
will have to  justify this anomalous negative correction on the
$\varepsilon_{N1}$ parameter. We show in this paper that a proper vector-family
could lower in a sizeable way the $\delta \rho$ parameter without affecting
$\varepsilon_{N3}$.

 \section*{The vector-like family} The model we study is a very
simple example of a vector-like family. As in SO(10) we consider a standard
fermion family plus a right-handed neutrino \cite{graham}, and we add
 a conjugate family
where the role of left-handed and right-handed spinors are inverted with
respect to the gauge interactions. We call
\be
(u_L,d_L), (\nu_L,e_L),u_R,d_R,\nu_R,e_R
\ee
the first standard family\footnote{Hereafter,
 we call standard family the usual
fermion family with the addition of a right-handed neutrino.} and \\
\be
(\tilde u_R,\tilde d_R), (\tilde \nu_R,\tilde e_R),\tilde u_L,
\tilde d_L,\tilde \nu_L,\tilde e_L.
\ee
\noindent the conjugate family.
Fermions within parentheses are SU(2) doublets
with  isospin \\ $(1/2,-1/2)$, while the others are singlets.

We allow only SU(2)$\times$U(1) invariant mass terms and  usual Yukawa
 interactions with a standard Higgs doublet, in order to have
(\ref{rho}) at the tree level.
Obviously it is possible to arrange the mass matrix between these fermions
in order to have positive corrections both to $\varepsilon_{N1}$ and
$\varepsilon_{N3}$. On the contrary we are interested to have negative
corrections
to $\varepsilon_{N1}$ without affecting $\varepsilon_{N3}$; in fact
in this  case the top mass upper bound is weakened.
We assume a SU(2)$\times$U(1) mass term between coloured particles
so large (compared with breaking terms)
that radiative corrections coming from this sector can safely be neglected.
We introduce this requirement to simplify our calculations.
Since every left-handed weyl spinor has its corresponding right-handed
counterpart we can construct four leptonic Dirac fermions, one SU(2)
doublet and two singlets, that we call
\ba
L&=&(\nu_L+\tilde \nu_R,e_L +\tilde e_R)\nonumber \\
\nu&=& \nu_R+\tilde \nu_L \\
e&=& e_R+\tilde e_L. \nonumber \\
\ea
Now we introduce a common invariant mass term also for the leptonic sector
\be
{\cal L}_{inv}=m_0 \bar L L + m_0 \bar {\nu}\nu
+ m_0 \bar e e
\ee
and
 the yukawa interaction which will introduce a SU(2)$\times$U(1) breaking
mass term is
\be \label{yukawa}
{\cal L}_{yuk}=\lambda_y (
 \bar L \phi \nu + \bar L \phi^c e +
\bar L \phi \nu^c +h.c.)
 \ee
where the $\nu^c$ is the charge conjugated of $\nu$.
After symmetry breaking the Higgs acquires a {\it vev} $(v_0,0)$
which introduces by (\ref{yukawa}) a mass matrix between fermions , which can
be parametrized in terms of only one parameter $m=\lambda_y v_0$.

If we choose a mass basis we obtain the physical masses and the
SU(2)$\times$U(1) gauge interactions among the mass eigenstates. Both
the lightest and the heaviest particle are neutral if $m < 2 m_0/3$
 , and a discrete symmetry
 makes the lightest neutral particle  stable\footnote{
Possible
implications to dark matter are beyond the scope of this paper.}.

 \section*{The radiative corrections of the vector-like family}
 We focus our attention, now, on
the radiative corrections induced by the model  described above. Following
the analysis described in \cite{ab}\cite{abc} we only have to compute
 two quantities
\ba
 e_1&=&{A_{33}(0) -A_{WW}(0) \over M_W^2}  \label{def}\\ e_3&=&{c\over s}
F_{30}(0)\nonumber \ea
where
 $\Gamma_{ij}=-i g_{\mu\nu} (A_{ij}(0)+q^2 F_{ij}(q^2)) +q_\mu q_\nu
(...)$ are the
two  point Green  functions (  one-particle  irreducible as  defined in
\cite{ab}) and $i,j=0,3,{\mbox{\scriptsize W}} $
are the SU(2)$\times$U(1)
indices of the gauge bosons. The $e_5$ contribution \cite{bcf} is
 negligible because we
consider large fermion masses (well above the threshold of production at LEP
I). We have
neither  vertex nor box  diagrams.  After the  computation of the vacuum
polarization of gauge bosons and taking the limit\footnote{This relation is not
necessary to make  $\delta\rho<0$ but we need it to make $\varepsilon_{N3}$
small.} $m<<m_0$ we obtain

\ba e_1&=& -{3 \sqrt{2}\over 10 \pi^2} G_F {m^4\over m_0^2} \\ e_3&=& {23\over
60}{G_F M_W^2\over 2 \sqrt{2} \pi^2} {m^2\over m_0^2}. \\  \nonumber \ea We
can see that, in the limit $m\rightarrow 0$ and $m_0$ constant (or equivalently
$m_0\rightarrow  \infty$ and $m$ constant) , our model gives no correction
because it corresponds to restore the $SU(2)\times U(1)$ invariance of the mass
matrix. As an example for the choice $m=400~GeV$ and $m_0=1000~GeV$
(with this
choice  the lightest particle has  $m_{light}=200~GeV$, outside the range
explorable by LEP II) we have the following numerical values
\ba e_1&=&-1.\times
10^{-2} \\ e_3&=&3.\times 10^{-4}. \\  \nonumber
\ea
It can be shown that if
$M=M_0+\delta M$ is the mass matrix among $n$ Dirac fermions\footnote{
The formula (\ref{formula}) can be generalized in the case of weyl fermions
simply adjusting the factor in front of the trace.}
$\psi=(\psi_1,...,\psi_n)$
 in a reducible or
irreducible vector-like representation ($n$-dimensional),
$M_0$ is a SU(2) invariant mass matrix
(which gives equal mass to all fermions, for instance $m_0$)
 and $\delta M$
is a SU(2) breaking matrix (with small elements compared with $m_0$,
introduced by a Higgs doublet acquiring
a {\it vev})
 we obtain the general formula for $\varepsilon_{N1}$
\be
\label{formula}
\varepsilon_{N1}= { \sqrt{2}\over 5 \pi^2} G_F~Tr~{\left(
\left[M^2,T_3\right]^2-\left[M^2,T_1\right]^2\right)\over m_0^2}+...
\ee
where the dots stand for terms which are suppressed by factors
of the type $\delta M^2_{i,j} /m_0^2$ which go to zero when $m_0\rightarrow
\infty$ and the symmetry breaking mass terms $\delta M_{i,j}$ remain constants;
 $T_3,T_1$ are the matrices of the gauge generators in our
representation ($T_3$ corresponds to $W_3$ and $T_1$ corresponds to $W_1$).
We can easily observe that, if
\be
\label{relat}
\left[M^2,T_3\right]= -\left[M^2,Y\right]= 0
\ee
holds, than  equation  (\ref{formula})  gives\footnote{We remind that the
commutator is anti-hermitian and its square is definite negative.}
$\varepsilon_{N1}>0$. So, the
particular  feature that in this case  allows a negative  contribution to
$\varepsilon_{N1}$ is the violation of (\ref{relat}). We remind the reader
that
we are considering the case of SU(2)$\times$U(1) breaking mediated by Higgs
doublets, which is strongly motivated by (\ref{rho}). In our simple model the
mixing in the squared mass matrix between  the $\nu_L$ and
$\tilde \nu_R^c $
fermions (with respectively $T_3=+1/2$ and $T_3=-1/2$) introduces in the
neutral
sector of the  representation a  violation of  equation  (\ref{relat}).
Alternatively one could introduce such violation in the charged sector but the
resulting pattern of charged particles is less interesting\footnote{ We have to
introduce for example fundamental particles with charge $>1$}

Therefore, in conclusion, let us suppose that  the top  quark is heavy
($M_{top}>150~GeV$) and that the $\varepsilon_{N1}$-value resulting at the end
of LEP I
program will remain small (for instance compatible with zero). Then we have to
introduce a negative contribution to this parameter . One solution could be
supersymmetry (through $e_5$) but if no particle shows up at LEP II then
an SO(10) vector family, with a precise choice of masses, might   be
 useful to solve that problem.
 Alternatively, if we believe
that equation (\ref{relat}) holds, we argue that in a perturbative theory (with
light Higgs) is difficult to evade the upper bound on the top mass coming
{}from $\varepsilon_{N1}$.
\vskip 1.5truecm
\noindent{\bf Acknowledgements}
\vskip .4truecm
I would like to thank R.Barbieri for very interesting and helpful discussions.
\vskip 1.5truecm
\pagebreak

\end{document}